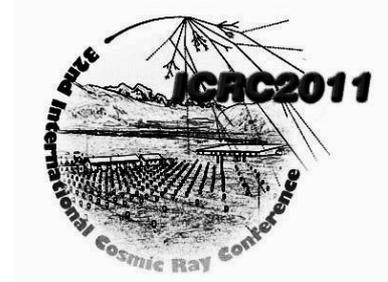

# Data Acquisition System for the UFFO Pathfinder

G. W. NA[1], K.-B. AHN[2], H. S. CHOI[3], Y. J. CHOI[4], B. GROSSAN[5], I. HERMANN[4], S. JEONG[1], A. JUNG[1], J. E. KIM[1], S.-W. KIM[2], Y. W. KIM[1], J. LEE[1], H. LIM[1], E. V. LINDER[1,5], K. W. MIN[4], J. W. NAM[1], K. H. NAM[1], M. I. PANASYUK[6], I. H. PARK[1], G. F. SMOOT[1,5], Y. D. SUH[4], S. SVERTILOV[6], N. VEDENKIN[6], I. YASHIN[6], M. H. ZHAO[1], FOR THE UFFO COLLABORATION

[1]*Ewha Womans University, Seoul, Korea*
[2]*Yonsei University, Seoul, Korea*
[3]*Korea Institute of Industrial Technologe, Ansan, Korea*
[4]*Korea Advanced Institute of Science and Technology, Deajeon, Korea*
[5]*University of California, Berkeley, USA*
[6]*Moscow State University, Moscow, Russia*
nagon82@gmail.com

**Abstract:** The Ultra-Fast Flash Observatory (UFFO) Pathfinder is a payload on the Russian *Lomonosov* satellite, scheduled to be launched in November 2011. The Observatory is designed to detect early UV/Optical photons from Gamma-Ray Bursts (GRBs). There are two telescopes and one main data acquisition system: the UFFO Burst Alert & Trigger Telescope (UBAT), the Slewing Mirror Telescope (SMT), and the UFFO Data Acquisition (UDAQ) system. The UDAQ controls and manages the operation and communication of each telescope, and is also in charge of the interface with the satellite. It will write the data taken by each telescope to the NOR flash memory and sends them to the satellite via the Bus-Interface system (BI). It also receives data from the satellite including the coordinates and time of an external trigger from another payload, and distributes them to two telescopes. These functions are implemented in field programmable gates arrays (FPGA) for low power consumption and fast processing without a microprocessor. The UDAQ architecture, control of the system, and data flow will be presented.

## 1 Introduction

The Ultra-Fast Flash Observatory (UFFO) is designed to study Gamma-Ray Bursts (GRBs). GRBs were first observed in 1967 by the U.S. Vela satellite, and they have been known as the most luminous electromagnetic event occurring in the universe. Since the first discovery, there have been many experiments to observe GRBs. *Swift*, launched in November 2004, is one of the successful space missions. It can be rapidly slewed to observe afterglow emission following a burst. However, it still takes about a minute as a response time to slew the spacecraft for the observation of GRBs; it is still not sufficient enough to study the initial emission of GRBs. Therefore, the UFFO is proposed to observe GRBs within seconds after a burst, and the main scientific goal of the mission is to detect the early UV/Optical photons from GRBs [1]. The UFFO pathfinder is the pilot model of the UFFO, and is scheduled to be launched in November 2011 on *Lomonosov* satellite in Russia.

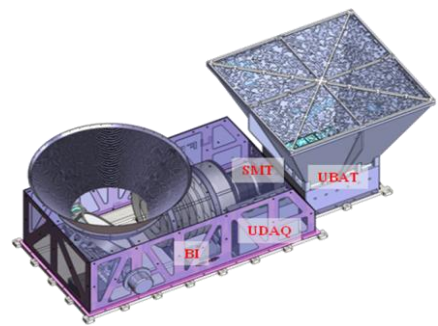

Figure 1: UFFO pathfinder

As shown in Figure 1, the UFFO pathfinder is comprised of two telescopes, the UFFO Burst Alert & Trigger Telescope (UBAT) and the Slewing Mirror Telescope (SMT), to detect the early UV/Optical photons from GRBs by fast trigger and a central data acquisition system, the UFFO Data Acquisition (UDAQ) system, to control the two telescopes. The UBAT detects the photons from the range of 5 to 200 keV of X-ray targets in the given field of view with a tungsten coded mask, Cerium-doped Lute-tium Yttrium Orthosilicates (LYSOs) crystal and 36 multi-anode photomultiplier tubes (MAPMTs), which each contains 64 channels. The half coded field of view (FOV) is 70.8° and the full coded FOV is 43.2°. The UBAT counts the number of photons per channel per 2.5μs and records the energy sum of 8 channels per 2.5μs [2]. Also, the SMT can detect the UV/Optical photons from 200 nm to 650 nm by using an intensified charge-coupled device (ICCD) [3]. It has a Ritchey-Chretien telescope and a motorized slewing mirror. The optics is designed to 11.4 F-number, 17 × 17 arc-minute FOV with 4 arc-second pixel resolution and 0.5 arc-second location accuracy. The slewing mirror uses a step motor and it can stabilize or target within 1 s [4].

This paper describes the design and the implementation of the data acquisition system and discusses the performance results.

## 2  The UFFO Data Acquisition System

### 2.1 Goals and purpose

The UDAQ controls and manages the operation and communication of each telescope, and is in charge of the interface with the satellite. Also the UDAQ will save the data taken by each telescope in the NOR flash memory. These procedures are implemented in the field programmable gates arrays (FPGA).

### 2.2 Design and fabrication

The requirement of the UDAQ hardware design for the UDAQ is as follows:

- power: 2 W
- weight: 170 g
- dimension: 160 × 130 mm$^2$
- optimization of the number/location of interface connectors
- low outgassing materials
- thermal dissipation
- space qualified components

The most critical factor for the UDAQ was to minimize the power consumption. This was accomplished by using Actel A3P1000 FPGA, for low power consumption and fast processing without a microprocessor. Therefore, we obtain less than 1 W power consumption. The dimension of the UDAQ is 160 × 130 mm$^2$ because it is located in the SMT case within the limited mechanical constraints such as the weight and the connections to systems. The UDAQ system is designed and fabricated using low outgassing materials, good thermal dissipation and space qualified components in order to get higher reliability for the space experiment. It is shown in Figure 2.

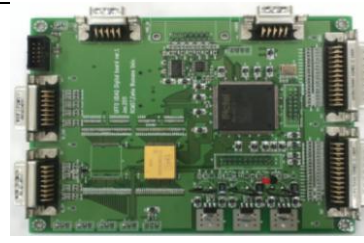

Figure 2: UFFO Data Acquisition (UDAQ) board

### 2.3 Architecture of the UFFO pathfinder

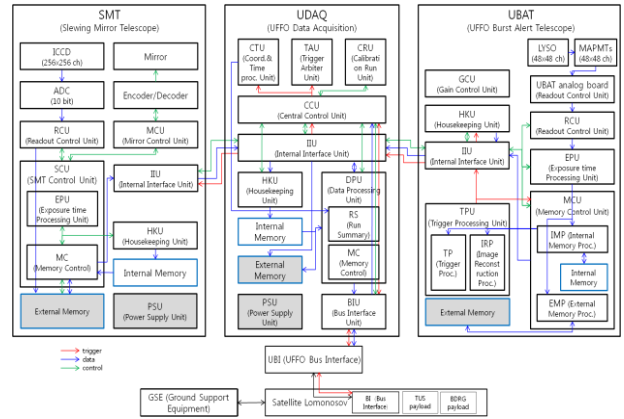

Figure 3: Architecture of the UFFO pathfinder

Figure 3 shows the architecture of the UFFO pathfinder for the trigger, the data, and the control. If the UFFO pathfinder is powered on, the UDAQ is operated by the auto mode sequence. The UDAQ collects the housekeeping information first, and gets the day/night information from the satellite. If the UDAQ figures out that it is dark enough to protect the ICCD from light, the UDAQ supplies the power to the SMT/UBAT by the power board, and all the control signals and parameters are distributed to the SMT/UBAT through the UDAQ to configure the UFFO pathfinder. If other configuration conditions are needed for other setups, commands are transferred to the UDAQ from the satellite via the Bus-Interface system (BI), and the UDAQ decodes and announces the commands to the applicable system. After the configuration is set, the UFFO pathfinder is ready to observe GRBs. The UBAT searches for GRB candidates by the trigger processing unit. If the trigger processing unit decides the UBAT detects a GRB candidate, the UDAQ gets a trigger signal and the position information of the triggered object from the UBAQ. The UDAQ then transfers them to the SMT and the SMT can observe the UV/Optical photons from the triggered object. Also, the UDAQ asks the BI to re-calculate the absolute coordinates of the triggered object and uses these coordinates information to detect the same object in the next orbit run. After taking the data by each telescope, all data are collected in the NOR flash memory in the UDAQ board. While the UDAQ collects the data, the UDAQ doesn't allow other control signals to run. Rejected or held control signals are resumed after the data collection without affecting the system. The UDAQ transfers the collected data to the satellite through the BI when the BI is ready to receive the data.

According to the main function, these above procedures can be classified into 3 categories: 1) communication, 2) data collection and transfer, and 3) system monitoring.

## 2.4 Communication

The communication of the UFFO pathfinder by the UDAQ is based on the customized serial peripheral interface bus (SPI). As Figure 4 shows, the UDAQ performs as a slave to communicate with the satellite via the BI. It receives the environmental conditions such as the coordinates, the time, even the external trigger from another payload, etc. and configuration commands from the satellite and it sends the collected data including its event header. On the other hand, the UDAQ communicates with each telescope as a master.

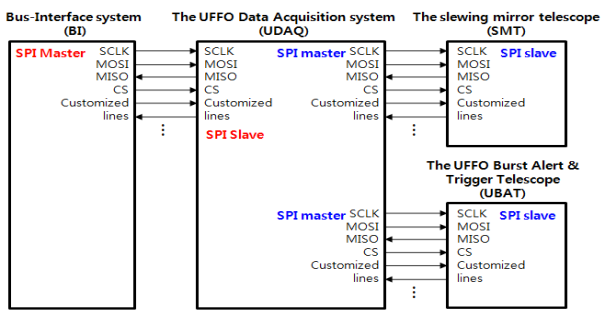

Figure 4: Diagram of the interface

There are 4 types of coordinates: the satellite coordinates, the BDRG coordinates, the UBAT absolute coordinates, and the UBAT relative coordinates. The position of the center of gravity of the satellite with respect to the earth is described by the satellite position coordinates, and the BDRG coordinates describe the direction of an external trigger source relative to the spacecraft-centered coordinate system. The UBAT relative coordinates describe the direction of a source detected by the UBAT relative to the UFFO pathfinder centered coordinate system. These coordinates information is transferred to the BI to calculate the absolute celestial coordinates corresponding to the triggered relative position by the UBAT, and these are the UBAT absolute coordinates. Each coordinate has 48 bits where the first 8 bits representing the coordinate indicator such as X, Y, Z, θ, ϕ etc., and 40 bits tell its value. The time information is composed of year, month, day, hour, minute, and second; each consists of 8 bit data. It is given with an accuracy of 1 second based on the satellite time. And a pulse is given to the UFFO pathfinder, every 1 minute to synchronize the real time clock between each system. Parameters such as the run mode, the run state, the trigger threshold, the configuration values, etc. are designed to be changed by commands. The command is forwarded to the UDAQ from the BI with 32 bits, and it is decoded and categorized by the UDAQ according to the command structure (Table 1). If the command is targeted to the SMT or the UBAT, the UDAQ sends the 64 bit command to the target. The 32 bit indicator is added to the 32 bit commands from BI. The BI allows for us to store 2,000 commands which can be used in the space mission. They are packed according to the run mode to operate the UFFO pathfinder in space by request from the ground.

| Content | Position (highest:lowest) | No. of digits | Example |
|---|---|---|---|
| Command header | 31:29 | 3 | single, packet |
| Applicable system | 28:24 | 5 | UDAQ, SMT, UBAT |
| Run type | 23:22 | 2 | calibration run, science run |
| Command content | 21:16 | 6 | state & transition, set, set parameter |
| Sub content | 15:10 | 6 | run type, trigger mode, etc. |
| Value | 9:0 | 10 | values |

Table 1: Command structure

These serial inputs – coordinates, time, and commands are stored in the serial-in/parallel-out register, and the parallel output is used in the UDAQ. At the same time the UDAQ converts it to serial output by using the parallel-in/serial-out register to send to each system.

## 2.5 Data collection and transfer

Event processing begins upon the receipt of a trigger signal from the UBAT or another external trigger system. The data taken from the SMT/UBAT is serialized to transfer to the UDAQ, and it is first temporarily stored serial-in/parallel-out register, implemented in FPGA. After then the 16 bit parallel-out data will be written to the NOR Flash memory within approximately 7 μs. The NOR flash memory has a 256 Mbit capability composed of 4 chips of 64 Mbit. One of chips will be used as a look-up table for the calculation for the coordinate and the triggering, and will also store the latest configuration parameters from the ground. The UDAQ saves the data in the remaining 3 chips organizing the addresses/chips according to event. The size of designed event is about 5 Mbyte; therefore, the UFFO pathfinder can save 2 events at once. The transmission speed from the UDAQ to the BI used to transfer to the satellite is 1Mbyte/s. The specification of the NOR flash memory is given as follows.

|  | NOR-256Mb |
|---|---|
| Write | 7 μs/16bit |
| Read | > 90ns |
| erase | - |
| Life time | 1000K write/erase |
| Radiation | 15Krads |

Table 2: Specification of the NOR flash memory

## 2.6 System monitoring

The system monitoring including the power control enables the UFFO pathfinder system to survive automatically. The 4 photo sensors, located in the SMT, are used to monitor the light intensity to protect the SMT detector system. The 10 temperature sensors are distributed inside the UFFO pathfinder case and monitor the temperature in each region. The power supplies of the SMT/UBAT (5.2 V and 12 V) are controlled by comparing the digital value of the currents monitored in the register with the threshold. The information monitored by each telescope is also stored in the UDAQ. If any of monitored values exceeds the threshold, it is written in the register and the UDAQ powers off the telescopes. And this is announced to the BI. If each telescope announces the emergency situation to the UDAQ, the UDAQ operates according to the manual of the situation.

# 3 Performance result

## 3.1 Development setup

The interface between the UDAQ and the SMT were successfully tested and demonstrated in spring 2011 as shown in Figure 5. The UDAQ could be controlled and send the data by using PC with USB8451. The user control panel is shown in Figure 6. Users can send the commands, the 4 types of the coordinates, and the time to the UDAQ to control the system. Also, they can receive the data from the UDAQ and find the law data in the panel. They can create the data file in the path they want, too. The panel simply shows the image from the data.

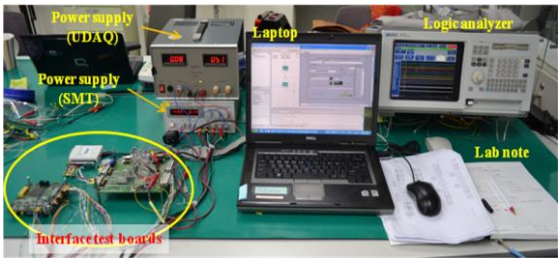

Figure 5: Interface test setup

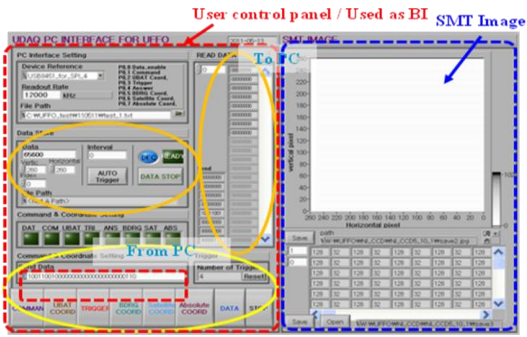

Figure 6: User control panel

## 3.2 Test result

The UDAQ system was first checked alone with its flags, which verify the task is performed or not. And then the UDAQ was set with the SMT to test the interface between them. Interface tests were checked to compare the signal from the SMT to the UDAQ by a logic analyzer (e.g., Figure 7) with the data by the PC. We got that all signals are correct. Therefore, we can confirm the UDAQ has a good performance for the main functions implemented in the FPGA.

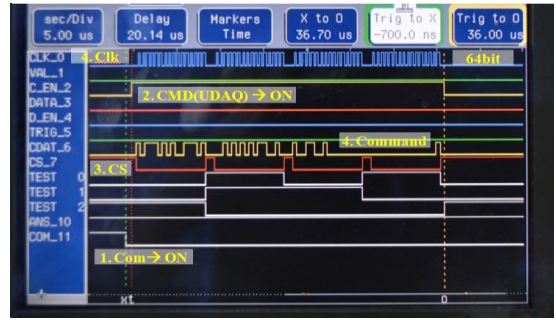

Figure 7: Command from the UDAQ to the SMT

# 4 Summary

The UFFO Data Acquisition (UFFO) system is designed to perform the main functions. The communication and the data collection and transfer, between the UDAQ and the SMT, were pre-tested successfully in the first quarter of this year. The UDAQ will be developed and verified until launching time.